# ASTRA Mission Concept Description

# LUnar-based Survey for Time-domain Exploration and Research network (LUSTER-net)

**Authors:** T. Boyajian (Louisiana State University)[1],
R.A. Street (Las Cumbres Observatory), B. Fleming (CU Boulder), F. Marchis (The SETI Institute), J. Morse (University of Arizona), P. Gandhi (University of Southampton), J. Turner (Cornell University), G. Rau (Catholic University of America), A. Tanner (Mississippi State University), A. Sohani (Louisiana State University)

**For ExoPAG, CoPAG, and PhysPAG**

[1] boyajian@lsu.edu

# 1. Science Investigation

LUSTER-net, the LUnar-based Survey for Time-domain Exploration and Research network, is a proposed lunar-surface UVOIR observatory network for time-domain astrophysics. The concept envisions a scalable array of ~6-12 telescope nodes distributed across the lunar surface to provide sustained access to transient and variable targets. Individual nodes would be in the ~0.5-1 m aperture class, a range that offers meaningful gains in collecting area and angular resolution while remaining compatible with commercially produced telescope systems and replicated-node development. Each node would provide imaging and/or spectroscopic capability from the ultraviolet through the optical and into the near-infrared, including UV bandpasses inaccessible from the ground. An ASTRA study would define the optimal number of nodes, lunar locations, aperture scale, instrument complement, and deployment cadence needed to maximize science return while leveraging commercial and Artemis-enabled lunar infrastructure. The central concept is a scalable lunar observatory network for high-cadence, long-duration monitoring of transient and variable sources from an exquisitely stable space-based platform.

Time-domain and multi-messenger astrophysics is entering a discovery-rich era. Rubin Observatory will generate an unprecedented optical alert stream; Roman will conduct high-cadence Galactic bulge time-domain observations and issue discovery alerts through RAPID; ULTRASAT and UVEX will expand wide-field ultraviolet time domain discovery; and ground-based surveys, gravitational-wave, neutrino, and high-energy facilities will continue to produce rapid transient alerts [e.g., 1-5]. These facilities will transform discovery, *but alerts alone do not provide the multi-wavelength light curves, spectra, astrometry, and long-term monitoring needed to determine the physical nature of events.* The community will need persistent, flexible, space-based UVOIR follow-up on timescales from minutes to months.

LUSTER-net would address this gap with a coordinated lunar network. The lunar surface provides long continuous observing windows without the repeated Earth-occultation interruptions typical of low-Earth orbit. The network architecture increases sky access, cadence, redundancy, target availability, and rapid-response capability relative to a single telescope. Imaging would provide light curves, colors, and astrometry, while spectroscopy would provide classification, redshift estimation, temperature/ejecta evolution, and physical diagnostics. Network-level intelligence would coordinate observations across nodes, optimize cadence and filter choices, respond to external alerts, and adapt to changing priorities. *LUSTER-net would not replace large ground-based surveys or future flagship observatories; it would provide the persistent UVOIR time-domain layer needed to convert discoveries into physical understanding.* LUSTER-net would also advance the broader lunar-astrophysics trajectory: a network of commonly designed, precisely calibrated, centrally scheduled lunar nodes would build the operational, timing, astrometric, and cross-node calibration heritage needed for more ambitious future lunar facilities, including surface interferometric arrays and broader wavelength coverage (e.g. X-ray, far-IR).

The science case is broad because the capability is broad. LUSTER-net would enable comparative exoplanet atmosphere studies through repeated UVOIR transit monitoring of diverse planetary and stellar environments [6,7]; microlensing parallax measurements from lunar, Earth-based, and space-based baselines to constrain lens masses and distances [8,9]; Solar System monitoring of NEA rotation, shape, binarity, composition, cometary dust/gas activity, outbursts, interstellar objects, and unusual small-body activity; early light curves, colors, and spectra of supernovae and fast transients to constrain progenitors, shock interaction, circumstellar material, and explosion physics [10]; rapid UVOIR follow-up of gravitational-wave, neutrino, and high-energy alerts to identify kilonovae, GRB afterglows, tidal disruption events, and other rare transients [11,12]; and broader UVOIR time-domain studies of accreting binaries, eruptive protostars, brown dwarf variability, stellar activity, and calibration standards.

LUSTER-net is aligned with Astro2020, the 2014 Astrophysics Roadmap, and the ASTRA initiative [13–15]. Astro2020 identifies time-domain and multi-messenger astrophysics as the highest-priority sustaining activity in space and emphasizes coordinated observations of transient phenomena [13]. The 2014 Roadmap frames NASA astrophysics around enduring questions including “How did we get here?” and “How does the universe work?” [14]. LUSTER-net addresses these questions by probing planetary atmospheres, lensing events, small bodies, stellar deaths, compact-object assembly, Galactic structure, accretion, explosions, and relativistic outflows.

LUSTER-net should enter the ASTRA incubator now because the discovery landscape and lunar implementation landscape are converging. Rubin, Roman, ULTRASAT, and multi-messenger facilities will sharply increase the number of transient alerts requiring rapid and sustained UVOIR characterization [1–5]. At the same time, NASA’s planned lunar infrastructure build-up, commercial lunar delivery, larger payload capacities, and improved communications make a staged lunar observatory network increasingly credible [16]. ASTRA is the appropriate venue to study whether this non-traditional architecture can provide strategic science return at a cost and risk profile competitive with conventional single-observatory designs [15].

The key science questions are: How do exoplanet atmospheric properties vary across planetary and stellar environments? Can lunar parallax break degeneracies in microlensing events and determine lens masses and distances, including for short-timescale and free-floating planet candidates? How can continuous UVOIR monitoring improve characterization of Solar System objects, including NEA physical properties and cometary dust/gas variability? What progenitor, explosion, and environment physics drive fast UVOIR transients? Which multi-messenger alerts produce detectable electromagnetic counterparts, and how do their UVOIR light curves and spectra reveal the underlying physical event? What broader classes of variable astrophysical sources can be characterized with a coordinated intelligent lunar UVOIR network?

## 2. Science Table

| Science Objectives | Physical Parameters | Observables | Potential Challenges |
|---|---|---|---|
| **Determine how exoplanet atmospheric properties vary across planetary and stellar environments.** | Atmospheric opacity, aerosols, escape signatures, stellar activity, irradiation, gravity, planet type. | Repeated UVOIR transits; full-transit baselines; color-dependent depths; epoch-to-epoch variability. | Requires high photometric stability; stellar contamination and atmospheric variability must be separated. |
| **Use lunar parallax to determine physical properties of microlensing lenses.** | Lens mass/distance, source distance, relative proper motion, planet/star mass ratio. | Earth/lunar microlensing light curves; parallax signal; event timescale; color/magnitude evolution. | Sensitivity depends on event geometry, cadence, and photometric precision; degeneracies remain for some events. |
| **Characterize solar-system objects and activity from continuous UVOIR monitoring.** | NEA orbit, rotation, shape, binarity, composition; comet dust/gas activity; interstellar-object activity; phase behavior; surface heterogeneity. | Astrometry; colors; time-resolved photometry; non-sidereal image sequences; comet/interstellar-object morphology; activity onset. | Requires non-sidereal tracking, ephemeris updates, rapid prioritization, edge processing, calibration, and selective downlink. |
| **Determine progenitor, explosion, and environment physics of fast UVOIR transients.** | Progenitor radius, ejecta mass/velocity, temperature, composition, circumstellar material. | Early multi-band light curves; color evolution; rise/decline rates; spectra; line velocities. | Requires rapid response, high cadence, spectroscopic sensitivity, and calibration across nodes. |
| **Identify and characterize EM counterparts to multi-messenger events.** | Merger ejecta mass, opacity, composition, viewing angle, jet energy, circumburst environment. | UVOIR counterpart detection; fading rate; color evolution; spectra; localization refinement. | Events are rare, fade quickly, and may have large localization regions; rapid and coordinated scheduling to alerts. |
| **Enable UVOIR monitoring of variable astrophysical sources with a calibrated lunar network.** | Accretion rates, outburst energetics, stellar/substellar atmospheric structure, solar activity, calibration stability. | UVOIR time series and spectra of accreting binaries, protostars, brown dwarfs, standard stars, and bright targets. | Broad use requires flexible scheduling, fast data transfer, stable calibration, bright-target handling, and community-access policies. |

# 3. Instrument Description

LUSTER-net would use replicated lunar UVOIR telescope nodes optimized for time-domain imaging and low-resolution spectroscopy. The node design should be studied over a plausible ~0.5-1 m aperture range, with wide-field imaging and broad-bandpass spectroscopy, such as a low-resolution long-slit mode from the ultraviolet through optical and potentially into the near-infrared. This range improves sensitivity and angular resolution for faint transients, microlensing, moving objects, and spectroscopic classification while remaining compatible with commercially produced telescope systems and replicated-node development. A common node design is preferred to simplify calibration, scheduling, software, operations, and staged deployment; final aperture, field of view, detector format, wavelength coverage, and spectral resolution should be refined through an ASTRA study.

Each node would include UVOIR-sensitive detectors, filter imaging, low-resolution spectroscopy, onboard calibration, aperture covers, onboard/edge computing, and standardized data products. Imaging would provide light curves, colors, astrometry, moving-object measurements, and host/environment context; spectroscopy would provide classification, redshift, line velocities, continuum evolution, and emission/absorption features that constrain composition, temperature, and physical state.

Onboard and edge processing should be treated as a core enabling capability. Each node could perform source detection, image differencing, moving-object detection, transient identification, data-quality assessment, event ranking, compression, and selective downlink. This would reduce communication burden, improve response time, and preserve the highest-value science products. Because downlink capacity could otherwise limit LUSTER-net to a narrower transient-follow-up role, an ASTRA study should evaluate data-volume mitigation strategies, including onboard event selection, prioritized image-stamp downlink, scheduled full-frame sampling, compression, and potential relay assets for broader survey-like operations.

Key technical drivers include photometric and spectrophotometric stability across nodes, pointing stability, near-UV throughput, detector radiation tolerance, thermal control, stray-light control, seismic calibrations, dust/contamination control, calibration transfer, and low-latency response to alerts. A deployable aperture cover would protect optical surfaces during descent, landing, and initial checkout. The Chang’e-3 Lunar-based Ultraviolet Telescope provides precedent for lunar UV astronomy, demonstrating stable photometric performance over months of surface operation [18].

LUSTER-net can reduce development risk by building on Earth-based robotic telescope-network experience, including adaptive scheduling, calibration pipelines, data processing, and community follow-up workflows [17]. Hardware must be adapted for the lunar environment, but the concept should use off-the-shelf or space-heritage components where possible. Replicated nodes, shared software, onboard processing, selective downlink, and staged deployment make this a scalable, lower-risk approach to a strategic UVOIR time-domain facility.

## 4. Mission Implementation

LUSTER-net is envisioned as a staged lunar-surface observatory network. An initial deployment would place one or a few nodes on the Moon to demonstrate UVOIR time-domain observing, calibration, alert response, and network operations; later deployments would expand toward 6-12 nodes to increase sky access, cadence, redundancy, wavelength coverage, and spectroscopy. The observing program would combine scheduled monitoring, rapid target-of-opportunity response, and observations coordinated with ground-based observatories.

The mission destination is the lunar surface, with site drivers including sky visibility, thermal environment, power, communications, dust/contamination risk, stray light, terrain, and lander accommodation. Unlike low-Earth orbit, the lunar surface provides long continuous observing windows, allowing a single node to monitor a target for multiple days with a stable calibration baseline rather than relying on frequent spacecraft handoffs that can introduce cross-calibration systematics. This enables uninterrupted light curves and sustained monitoring from minutes to weeks. The mission lifetime is TBD; an initial deployment could demonstrate one lunar day, while the strategic goal is repeated lunar-day operations enabled by survive-the-night capability and eventual night operations.

The architecture requires landed nodes with UVOIR imaging and/or spectroscopy, onboard processing, calibration systems, aperture covers, and efficient downlink of image stamps, spectra, alerts, and selected larger fields. A key design feature is an adaptive, centrally coordinated scheduler that responds to alerts, coordinates lunar and ground-based observations, and adjusts cadence, pointing, and target priority for evolving or moving objects. ASTRA trades include node number, with 6-12 as an initial scale; aperture, with ~0.5-1 m as a commercially plausible range; site distribution; imaging/spectroscopy balance; response time; data volume; lander-provided services; non-sidereal tracking; and node commonality.

LUSTER-net is well matched to ASTRA's emphasis on non-traditional architectures and partnerships [15]. CLPS and NASA's planned lunar infrastructure provide pathways for staged deployment, lander accommodation, power, communications, and larger payload capacities [16,20]. Potential partnerships include industry for lander services, modular payload development, and autonomous operations; universities and NASA centers for instrumentation, calibration, scheduling, science operations, and analysis; and international partners for additional nodes, ground support, complementary follow-up, data processing, computing, and infrastructure, building on models such as the Rubin In Kind program [21].

Cost-saving features include replicated node designs, staged deployment, commercial lunar delivery, limited moving parts, off-the-shelf or space-heritage components, selective downlink, and reuse of open-source telescope-network software and Earth-based robotic-network experience. These choices reduce mission-unique software development for scheduling, calibration, and data processing. The network can grow incrementally: early nodes provide pathfinder science and retire implementation risks, while later nodes expand LUSTER-net into a strategic UVOIR time-domain observatory.